\documentclass[prl,twocolumn,english]{revtex4-1}

\usepackage{textcomp}
\usepackage{amstext}
\usepackage{graphicx}
\usepackage{amsmath}
\usepackage{float}
\usepackage{verbatim}   % useful for program listings
\usepackage{color}      % use if color is used in text
\usepackage[colorlinks=true, allcolors=blue]{hyperref}
\usepackage[all]{hypcap}

\begin{document}
\title{Non-Monotonic Aging and Memory retention in Disordered
Mechanical Systems}

\author{Yoav Lahini, Omer Gottesman, Ariel Amir and Shmuel M. Rubinstein}

\affiliation{Harvard John A. Paulson School of Engineering and Applied Sciences, Harvard University, Cambridge, Massachusetts 02138, USA }

\begin{abstract}
		We observe non-monotonic aging and memory effects, two hallmarks of glassy dynamics, in two disordered mechanical systems: crumpled thin sheets and elastic foams. Under fixed compression, both systems exhibit monotonic non-exponential relaxation. However, when after a certain waiting time the compression is partially reduced, both systems exhibit a non-monotonic response: the normal force first increases over many minutes or even hours until reaching a peak value, and only then relaxation is resumed. The peak-time scales linearly with the waiting time, indicating that these systems retain long-lasting memory of previous conditions. Our results and the measured scaling relations are in good agreement with a theoretical model recently used to describe observations of monotonic aging in several glassy systems, suggesting that the non-monotonic behavior may be generic and that a-thermal systems can show genuine glassy behavior.
\end{abstract}

\maketitle
Many disordered systems exhibit phenomenologically similar slow relaxation
dynamics that may span many time scales - from fractions of a second
to days and even years. Examples range from time-dependent resistivity
in disordered conductors \cite{grenet2007anomalous, Vaknin2000,pollak2013electron, Amir2011Huge,ArielAmirAnuualReview}, flux creep in superconductors \cite{Kim1962,Anderson1962}, dynamics of spin glasses \cite{cugliandolo1993analytical,bouchaud1998spin,castillo2002heterogeneous,dupuis2005aging}, 
structural relaxation of colloidal glasses \cite{weeks2000three, cipelletti2011glassy}, time-dependence of
the static coefficient of friction \cite{Rubinstein2006, rubinstein2009, Ben-David2010}, thermal expansion of polymers \cite{Kovacs1963,Struik1977}, compaction in agitated granular systems \cite{Knight1995}, and crumpling of thin sheets under load \cite{Albuquerque2002,Matan2002}.
The ubiquity of slow relaxation phenomena suggests the existence of
common underlying physical principles \cite{palmer1984models,sibani1989hierarchical, bouchaud1992weak,bouchaud1998out,bouchaud2000aging,Kolvin2012,Amir2012OnRelaxation}. However, as slow relaxation is usually a smooth featureless process, it is hard to discern between the different descriptions using experiments. One way of probing deeper into the time dependent properties of glassy systems is using a phenomenon known as aging, where the manner in which the system relaxes towards equlibrium depends on its history.

\begin{figure} [h]
	 
	\includegraphics[clip,width=3.3in]{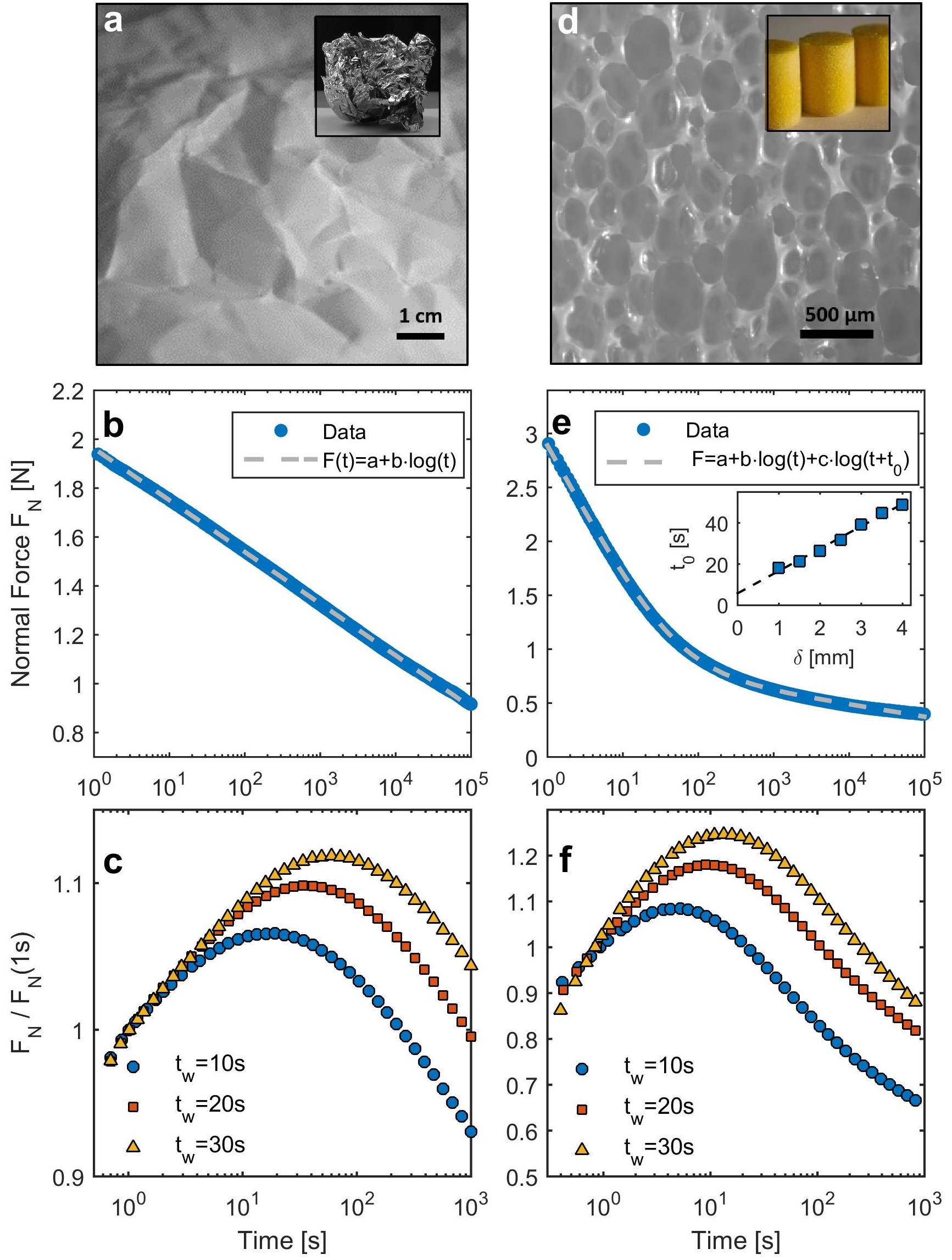}

	\caption{Two disordered mechanical systems. (a) The network of creases decorating a crumpled sheet of Mylar. Inset: a crumpled Mylar ball. (b) Stress relaxation of the crumpled ball with $H_1=45mm$, $\delta=H_1-H_2=5mm$. The dashed line is a fit to a logarithmic decay. (c) Non-monotonic  relaxation of crumpled Mylar, initially compressed by $\delta=5mm$ for $t_w$, and then released by $\Delta=2mm$. (d) Typical microscopic image of the cross-section of an PVC foam. Inset: elastic foam samples. (e) Stress relaxation of elastic foam with $H_1=18mm$ and $\delta=4mm$. The dashed line is a fit to $F=a+b\cdot log(t)+c\cdot log(t+t_0)$. Inset: $t_0$ vs $\delta$. (f) Non-monotonic relaxation for elastic foam with $\delta=3mm$ and $\Delta=1.5mm$.  \label{fig: 1} }

\end{figure}

\begin{figure}[b]
	\includegraphics[clip,width=3.4in]{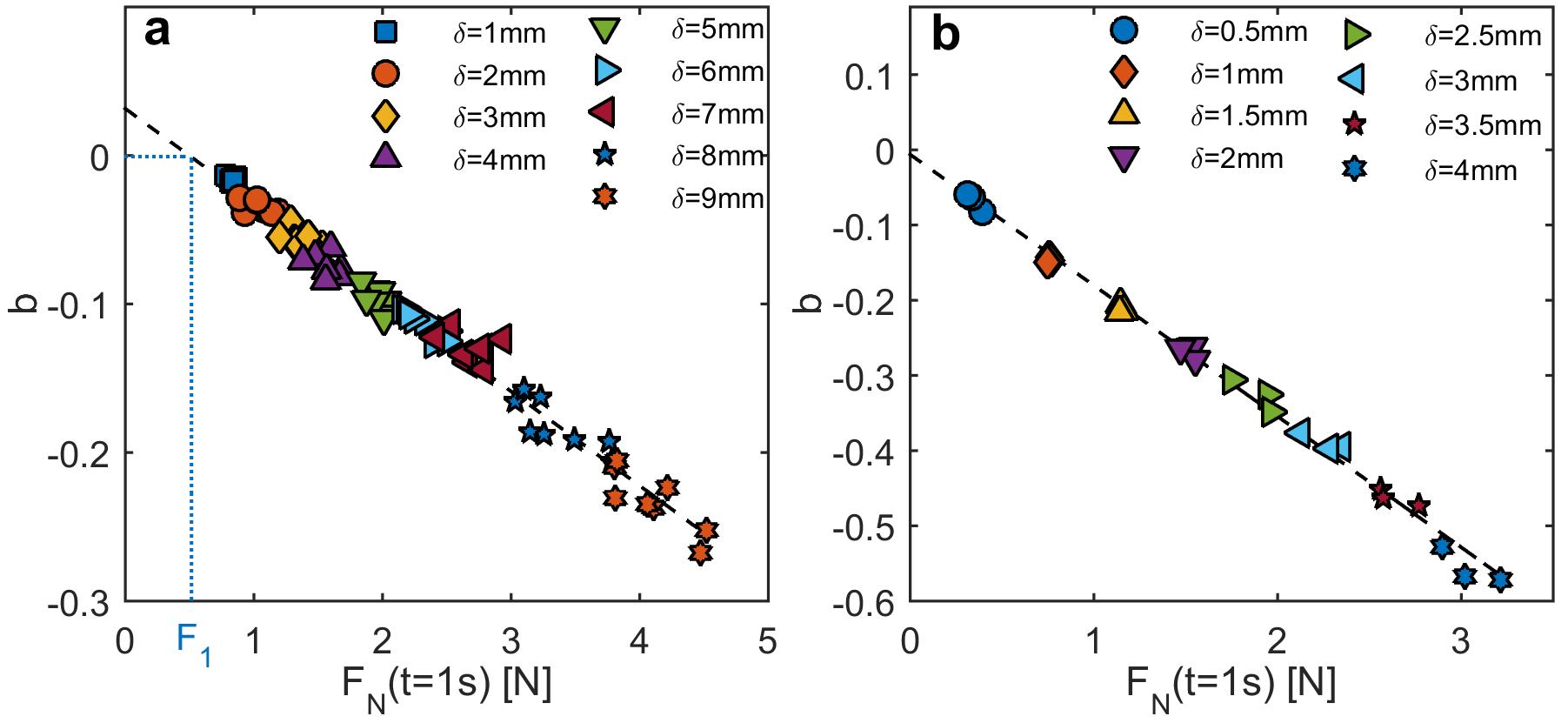}
	\caption{Reproducible stress relaxation. (a) crumpled Mylar: The relaxation rate $b$ vs the normal force at $t=1s$, $F(t=1s)=a$. The different symbols represent experiments preformed with different values of $\delta$. The normal force for which $b=0$ is marked as $F1$ (dotted blue line). (b) Same as (a)  for elastic foams, where here $F_N(t=1s)=a+c\cdot log(t_0+1)$ and we fit each relaxation curve to $F_N (t)=a+b\cdot log⁡(t)+c\cdot log⁡(t+t_0)$.  \label{fig: 2}}
\end{figure}
In this Letter, we report \textit{non-monotonic} aging dynamics that
give rise to a maximum in the relaxation curve. This extremum provides an
unambiguous signature of aging and memory, as well as a clear, measurable
time-scale. We experimentally study two distinct disordered mechanical
systems: crumpled thin sheets and elastic foams, shown in Fig. \ref{fig: 1}. When compressed, both systems exhibit monotonic, slow stress relaxation (Fig. \ref{fig: 1}b,e). When the compression is decreased after a certain waiting time, the stress evolution remarkably becomes non-monotonic: under constant compression,
the measured normal force first increases slowly over seconds to hours,
reaches a well-defined peak, and then reverses to a renewed slow relaxation (Fig. \ref{fig: 1}c,f). In both systems, the stress peak-time is linear in the waiting time,
indicating that the different systems carry a similar, long-lasting
memory of previous mechanical states. These observations are inconsistent
with the single-parameter model used to
explain logarithmic relaxation in crumpled sheets \cite{Matan2002},
yet are in agreement with a different phenomenological framework, successfully
used recently to define a new universality class related to the generic
behavior of aging in several glassy systems \cite{Amir2012OnRelaxation}.

Slow relaxation and aging experiments are performed in a custom
uniaxial compression tester. Samples are compressed between two parallel
plates, separated by a gap, $H$, which is set by a motorized stage. The compressive normal force, $F_{N}$, is monitored using an S-beam load cell (Futek LSB200) acquired at 24 kHz. We measure the stress relaxation behavior of thin Mylar sheets, 33cm\texttimes 33cm\texttimes 15$\mu m$, crumpled into
a ball, as shown in Fig. \ref{fig: 1}a. Samples are placed between
the plates of the apparatus, separated by an initial gap of $H_{1}$.
The gap is then reduced to $H_{2}<H_{1}$ and is held constant for the
rest of the experiment. Under these conditions the crumpled
sheets exhibit logarithmic stress relaxation, as shown for a typical
example in Fig. \ref{fig: 1}b. Such behavior was observed by Matan et.
al. \cite{Matan2002} and later by others \cite{Balankin2011}. Similar slow relaxations spanning several decades in time are exhibited by samples of elastic foam: dense open-cell porous materials made of elastic PVC, 18mm in height and 10mm in diameter, shown in Fig. \ref{fig: 1}d.

We perform a comprehensive set of stress relaxation tests on both
materials, keeping $H_{1}$ constant and measuring relaxation
curves for different compressions $\delta=H_{1}-H_{2}$. For crumpled Mylar we quantify the relaxation by fitting the curve to $F_{N}(t)=a+b\cdot log(t)$. Here, $a$ is related to the normal
force measured one second after the compression and $b$ is the logarithmic
relaxation rate. Typically, for larger compression steps
both $a$ and $b$ are larger. However, as reported in previous work on
relaxation in crumpled sheets \cite{Matan2002}, we find that the
relaxation curves fluctuate strongly between runs and no systematic
relation appears between $\delta$, $a$ and $b$. This irreproducibility
hampers any attempt to quantify the slow relaxation and the more subtle
aging behavior reported below. To this end, we identify an experimental
procedure in which the randomly crumpled sheets are "trained" before
the experiments, and as a result yield reproducible
behavior. First, new sheets of Mylar are repeatedly crumpled, opened and flattened. To achieve the same maximum compaction during training, the sheets were always crumpled into a cylinder, 65mm in diameter and 35mm in height. After at least 30 iterations,
additional crumpling of the sheet creates very few new creases
\cite{Gottesman}. Second, before each experiment we perform a quick
compression and release of the crumpled ball.  Finally, $H_1=45mm$ is used for the crumpled sheets, as we found it to be the maximal gap beyond which the crumpled ball could slip out of the apparatus. the normal force measured after training, which we denote as $F_1$, is approximately 0.5 Newton in all experiments. The corresponding relaxation rate, however, is negligible (see Figure \ref{fig: 2}a). The elastic foams do not require any support, thus here $H_1$ is the height of the sample. The value for maximum compression $H_2=5mm$ for the crumpled sheet and $H_2=3mm$ for the elastic foams, was chosen such that the experiments remain in the linear strain-stress regime. For higher compressions we observed a transition in to a power-law dependence \cite{Matan2002, Vliegenthart2006,Bai,deboeuf2013comparative}. Under these conditions we observe reproducible logarithmic relaxation curves, as shown in Fig. \ref{fig: 2}a. In particular we find a linear relation between $b$ and $F(1s)$ that is offset by $F_1$. The elastic foams require no training; measurements are reproducible as long as the sample is allowed to relax back to its original state between tests. Here the relaxation curves for all compressions $\delta$ can be fitted to a double-logarithmic function of the form $F_{N}=a+b\cdot log(t)+c\cdot log(t+t_{0})$. $b$ and $c$ are proportional to $\delta$ while the ratio between them remains approximately constant over all the relaxation curves. The inherent time scale $t_{0}$ shows a linear dependence on $\delta$, as shown in the inset of Fig. 1e.

The reproducible relation between the compression $\delta$
and the relaxation rate enables a systematic investigation of
the more subtle aging and memory effects which are observed after a sequence of compressions. 
Usually, the notion of aging implicitly assumes a slow monotonic process; 
however, in both systems we find that a two-step compression protocol results in non-monotonic aging dynamics and memory effects. 
Here, a sample is placed
between the two plates of the apparatus, separated by a gap $H_{1}$;
the gap is then decreased to $H_{2}<H_{1}$, and held constant for
a specific waiting time, $t_{w}$. During this first step, the normal force
monotonically decreases. At $t=t_{w}$, the gap between the
plates is increased to $H_{3}$ such that $H_{2}<H_{3}<H_{1}$,
and held constant for rest of the experiment. The subsequent dynamics separate into three distinct stages.
First, during the gap increase from $H_{2}$ to $H_{3}$, $F_{N}$ shifts abruptly to a lower value due to an elastic
response of the samples. Subsequently, in contrast to the naive expectation
that $F_{N}$ should now decrease at a logarithmic
rate that corresponds to the new compression, the normal force exhibits
a slow, non-monotonic behavior. Under constant external conditions, $F_{N}$
first slowly increases over many minutes and even hours. $F_{N}$
reaches a well-defined force peak at a time $t_{p}$, after which, for $t>t_{p}$, $F_{N}$
crosses back to a slow decay. Very similar non-monotonic
dynamics are measured for both systems, as shown for crumpled
Mylar sheets in Fig. \ref{fig: 1}c and elastic foams in Fig. \ref{fig: 1}f. We note that these systems also show non-monotonic volume relaxation when subjected to a two step loading protocol (not shown).

At any two time points in the non-monotonic dynamics in which $F_{N}$ has the same value before and after the peak, the sample's compression and all other macroscopic observables are identical; however, the system's evolution at these two points
is qualitatively different. Thus, the non-monotonic behavior clearly
indicates that the state of the system cannot be described by
the macroscopic observables alone, and additional degrees
of freedom storing a memory of the system's history must exist.

\begin{figure}
	\includegraphics[clip,width=3.4in]{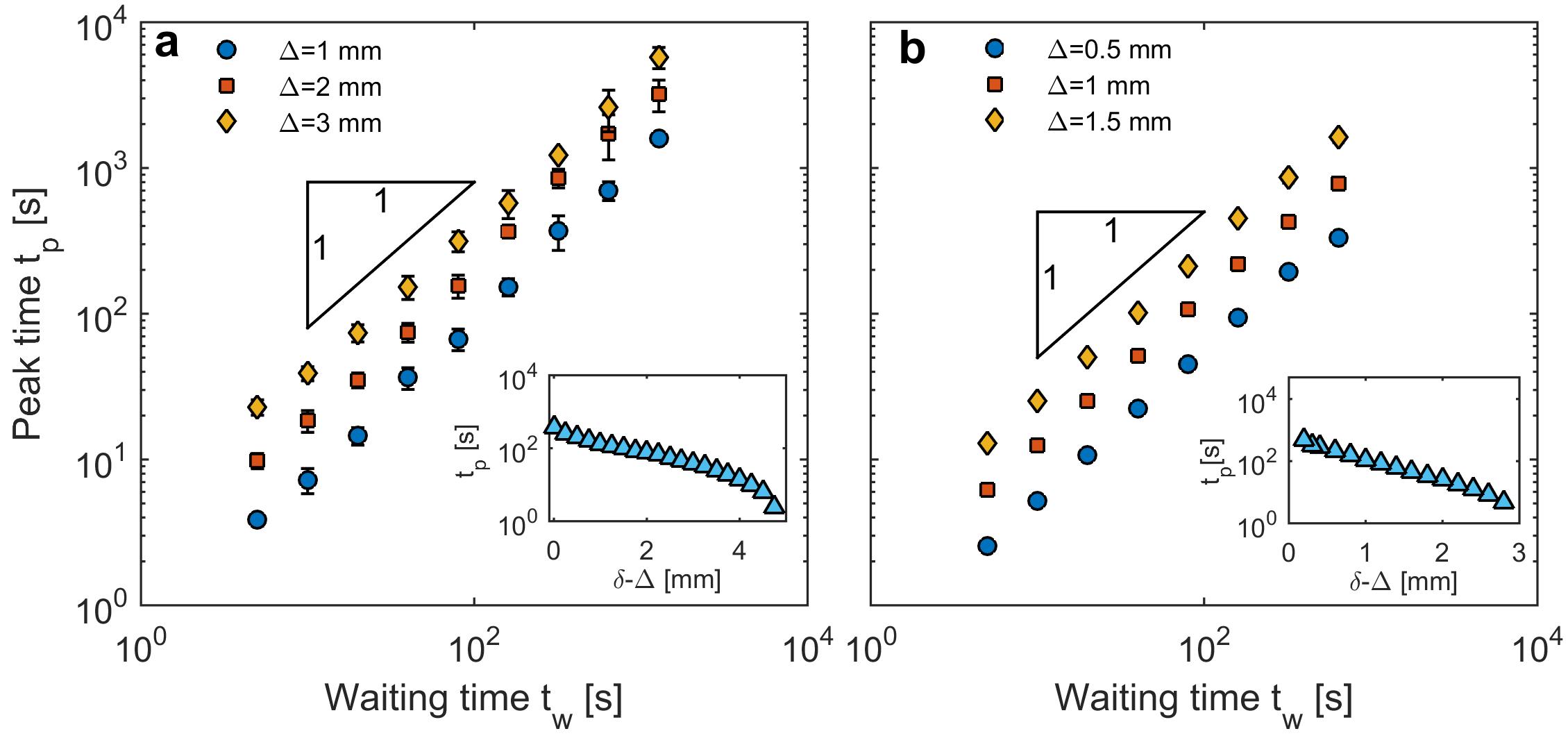}
	\caption{Memory effect. Linear scaling between the peak time and the waiting time for different values of $\Delta$, shown for crumpled thin sheets (a), and for elastic foams (b). Insets: peak time vs $H_1-H_3$  for $t_w=20s$. \label{fig: 3}}
\end{figure}

To characterize this non-monotonic behavior, we performed a systematic
study of the relation between the peak time $t_{p}$, the waiting time $t_{w}$  and the change in compression at the last stage $\Delta=H_{3}-H_{2}$. Here,
the reproducibility of the experiments is crucial. For fixed values
of $\Delta$, the relation between the waiting time $t_{w}$ and the
peak time $t_{p}$ is approximately linear over several decades. Increasing $\Delta$ results in a steeper linear dependence.
These results are depicted in Fig. \ref{fig: 3}a for the crumpled sheets and
Fig. \ref{fig: 3}b for elastic foams. Additional measurements in which $t_{w}$
was kept constant while $\Delta$ was varied over a wide range indicate
that the peak time increases as $H_{3}$ approaches $H_{1}$, as shown
in the insets of Fig. \ref{fig: 3}a and \ref{fig: 3}b. 

\begin{figure}[b]
	\includegraphics[clip,width=3.3in]{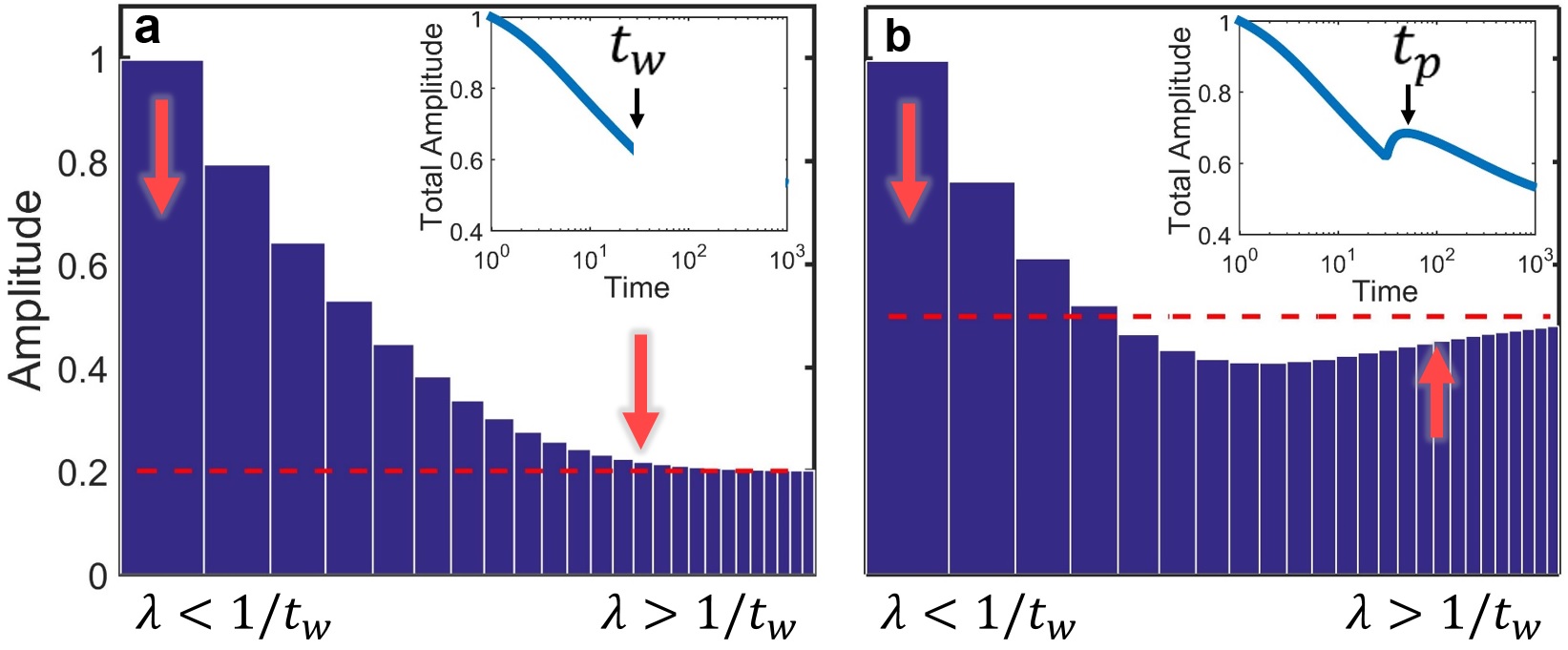}
	\caption{Phenomenological model. Simulation of Eq (3) for $V_1^{eq}=1$; $V_2^{eq}=0.2$, $V_3^{eq}=0.5$ and $t_w=30$. The instantaneous amplitude of the relaxation modes are shown for: (a) $t=t_w$ and (b) $t=t_p$. The slow modes with $\lambda > 1/t_w$ and the fast modes with $\lambda < 1/t_w$ are depicted on the left and right respectively . The width of the bars represents the abundance of the different relaxation times according to $P(\lambda)\propto 1/\lambda$. (Insets) Total amplitude as a function of time calculated by summing over all the individual modes. \label{fig: 4}}
\end{figure}

The scaling between
$t_{w}$ and $t_{p}$ is a hallmark of a memory effect - the time
in which the system reached its peak normal force is correlated with
changes in external conditions made up to several hours earlier.
These observations rule out single degree of freedom
descriptions previously suggested to model slow relaxations
in several disordered systems \cite{Anderson1962,Caroli2006}, including
crumpled thin sheets \cite{Matan2002}. Single-parameter theories
relate the relaxation rate of some macroscopic observable to its instantaneous value 
and thus cannot account for non-monotonic behavior, or for history
dependent evolution - i.e. memory. An alternative phenomenological
framework was recently used successfully to describe aging in several glassy
systems, introducing a new universality class related to the generic
behavior of logarithmic aging \cite{Amir2012OnRelaxation}. We show
that this framework can be generalized to apply also to
the experiments discussed here, capturing both the non-monotonic
relaxation, as well as the observed linear scaling
between $t_{p}$ and $t_{w}$. We assume a system which is controlled by a single parameter, $E$,
and which evolves via an ensemble of independent
exponential relaxation modes, each characterized by a rate
$\lambda$ with a broad distribution of rates, $P(\lambda)$. A key
assumption is that for every $E$ there exist an equilibrium state
$V^{eq}$ and that all relaxation modes have the same amplitude and
thus contribute to it equally. If a system is initially at the equilibrium
state $V_{1}^{eq}$ when $E$ is switched to a different value, its
relaxation towards a new equilibrium $V_{2}^{eq}$ can be written
as %$V(t)=V_{2}^{eq}+(V_{1}^{eq}-V_{2}^{eq})\intop_{\lambda_{min}}^{\lambda_{max}}P(\lambda)e^{-\lambda t}d\lambda$, 
\begin{equation}
	V(t)=V_{2}^{eq}+(V_{1}^{eq}-V_{2}^{eq})\intop_{\lambda_{min}}^{\lambda_{max}}P(\lambda)e^{-\lambda t}d\lambda\label{eq: 1}
\end{equation}
where $\lambda_{min}$ and $\lambda_{max}$ are physical cutoff
rates. Specifically, for $P(\lambda) \propto 1/\lambda$ and $1/\lambda_{max} \ll t \ll \ 1/\lambda_{min}$
we recover the logarithmic relaxation observed for the crumpled balls: $V(t)=V_2^{eq}-(V_1^{eq}-V_2^{eq})(\gamma_E+log(\lambda_{min}t))\equiv a+b\cdot log(t)$, where $\gamma_{E}$ is the Euler-Mascheroni constant. This particular distribution was shown to arise in certain disordered systems via several potential mechanisms \cite{Amir2012OnRelaxation}, including thermal activation (also leading to 1/f noise \cite{van1950noise,amir20091}) and multiplicative processes. Interestingly, the same distribution also arises in the context of "sloppy modes" \cite{waterfall2006sloppy}, and random matrices \cite{cao2016genuine,beenakker1997random}.

This formalism can predict the observed non-monotonic relaxations, without additional
assumptions. Here, starting at equilibrium $V_{1}^{eq}$, the system
evolves towards $V_{2}^{eq}$ only for a finite
time $t_{w}$ -- as shown schematically in Fig. \ref{fig: 4}a. At this point, the equilibrium
state shifts to $V_{3}^{eq}$. If $V_{1}^{eq}>V_{3}^{eq}>V_{2}^{eq}$
then at $t=t_{w}$ different modes can be found at different sides
of the equilibrium, as shown Fig. \ref{fig: 4}b. The slow modes, with decay rate $\lambda\ll1/t_{w}$ , are still
in the vicinity of $V_{1}^{eq}$, i.e. above $V_{3}^{eq}$, while
the fast modes with $\lambda\gg1/t_{w}$ have reached the new equilibrium
$V_{2}^{eq}$, and are below $V_{3}^{eq}$. Thus, immediately
after $t_{w}$ the dynamics of the fast and slow modes are in opposite
directions. At this stage the overall response can be dominated by
the fast modes and as a result $V(t)$ increases over time (Fig. \ref{fig: 4}b). After the fast modes reach the new equilibrium,
the overall response is dominated by the slow modes, leading to resumed relaxation.

\begin{figure}
		\includegraphics[clip,width=3.4in]{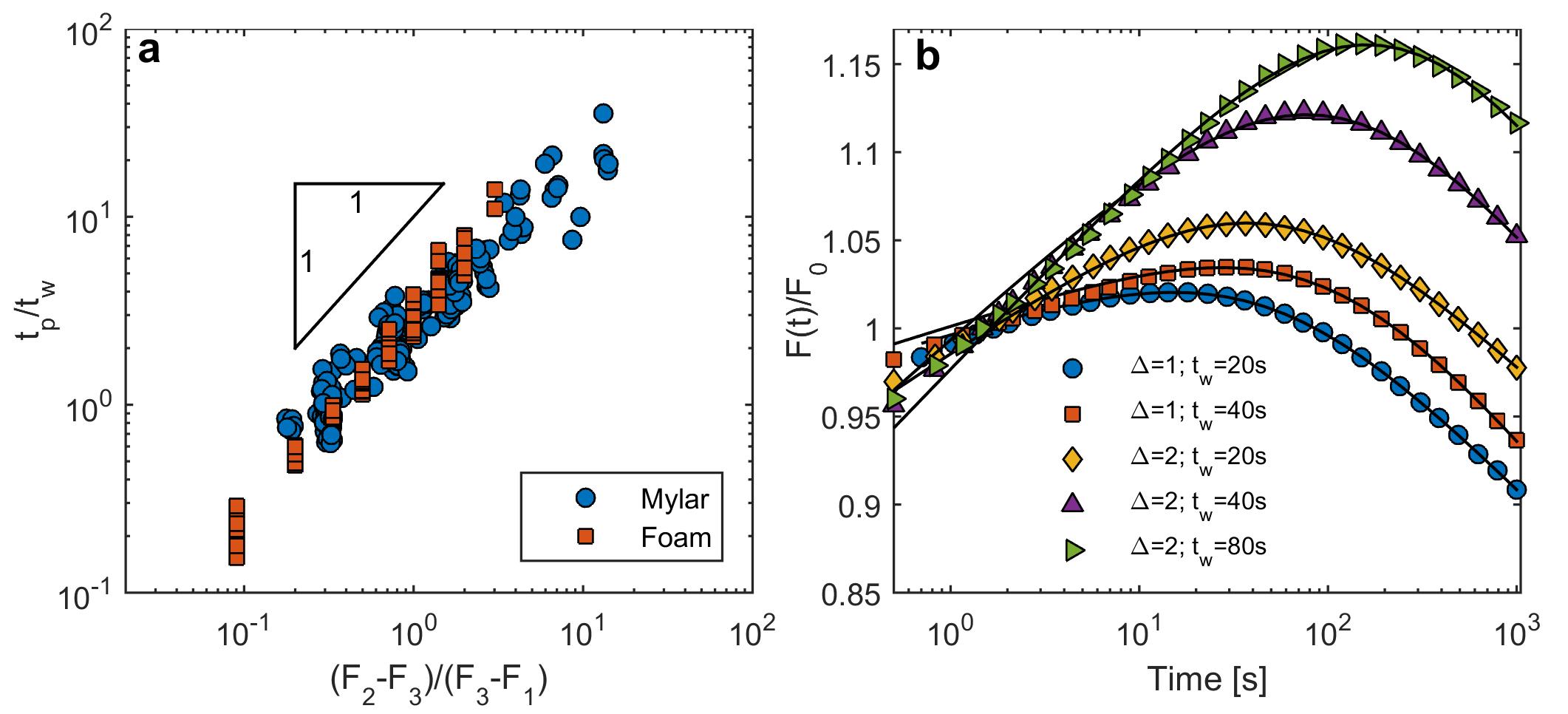}
	\caption{Universal relaxation dynamics. (a) $t_{p}/t_{w}$ versus $(F_{2}-F_{3})/(F_{3}-F_{1})$ for all the experiments preformed on crumpled sheets (blue circles) and elastic foams (red squares). (b) Single parameter fits for the non-monotonic relaxation of Mylar sheets to $A-B[(F_3-F_2)log(t)+(F_2-F_1)log(t+Ct_w)]$ with $B=A*0.025$ and $C=2.55$. \label{fig: 5}}
\end{figure}

Eq. 1 can be generalized for multiple steps by accounting for the out-of-equilibrium state
of each mode at time $t_{w}$. At this time, the state of each relaxation mode is given by 
$V_{3,\lambda}(t)=V_{3,\lambda}^{eq}+(V_{2,\lambda}(t_{w})-V_{3,\lambda}^{eq})\cdot e^{-\lambda t}$
with $V_{2,\lambda}(t_{w})=V_{2,\lambda}^{eq}+(V_{1,\lambda}^{eq}-V_{2,\lambda}^{eq})\cdot e^{-\lambda t_{w}}$.
Thus, for $t>t_{w}$ the system's evolution is given by %$V_{3}(t)=V_{3}^{eq}+(V_{2}^{eq}-V_{3}^{eq})\int_{\lambda_{min}}^{\lambda_{max}}P(\lambda)\cdot e^{-\lambda t}d\lambda +(V_{1}^{eq}-V_{2}^{eq})\int_{\lambda_{min}}^{\lambda_{max}}P(\lambda)\cdot e^{-\lambda(t+t_{w})}d\lambda$.
\begin{multline}
V_{3}(t)=V_{3}^{eq}+(V_{2}^{eq}-V_{3}^{eq})\int_{\lambda_{min}}^{\lambda_{max}}P(\lambda)\cdot e^{-\lambda t}d\lambda \\ +(V_{1}^{eq}-V_{2}^{eq})\int_{\lambda_{min}}^{\lambda_{max}}P(\lambda)\cdot e^{-\lambda(t+t_{w})}d\lambda\label{eq:3}
\end{multline}
As before this expressions can be approximated by:
\begin{multline}
V_{3}(t)=V_{3}^{eq}-(V_{2}^{eq}-V_{3}^{eq})(\gamma_{E}+log(\lambda_{min}t))\\-(V_{1}^{eq}-V_{2}^{eq})(\gamma_{E}+log(\lambda_{min}(t+t_{w})))\label{eq:4}
\end{multline}
Turning back to the experiments, the equilibria values $V^{eq}$
represent the normal forces as would be measured at infinitely long
time, where the equilibrium force is larger for lower $V^{eq}$. However,
this regime is not attainable experimentally as the normal forces
we measure do not show any signs of reaching equilibrium. Nevertheless, according to the model, $b$ is proportional to changes in the $V^{eq}$, so that $V^{eq}_2-V^{eq}_3 \propto b_2-b_3$ etc. As shown in Figure \ref{fig: 2}a, for the crumpled Mylar sheets $b$ is proportional to $F(t=1s)-F_1$ Thus we can replace $b_i$ with $F_{i}-F_1=F(H_{i}),(i\in{2,3})$, where $F_{i}$ is the normal force as measured one second after a compression from $H_{1}$ to $H_{i}$, and $F_1$ is the normal force measured at $H_1$ just before the compression. Using this substitution
and by differentiating Eq. 3 to find the curve maximum, we find:
\begin{equation}
t_{p}/t_{w}=(F_{2}-F_{3})/(F_{3}-F_{1})\label{eq:5}
\end{equation}
Using this scaling relation, the data from all experiments
performed on crumpled Mylar approximately collapses to a single linear curve as shown in Fig. \ref{fig: 5}a.
The analysis reveals an additional constant, denoted $C$, such that the collapsed curve is of the form
$t_{p}/t_{w}=C(F_{2}-F_{3})/(F_{3}-F_{1})$ with $C=2.6\pm0.2$. Accordingly, it is possible to fit the non-monotonic relaxation curves to a modified version Eq. 3, namely: $V_{3}(t)=A+B\cdot[(F_{3}-F_{2})\cdot log(t)+(F_{2}-F_{1})\cdot log(t+Ct_{w})]$, using $A$ as a fitting parameter and with $B=A*0.025$ and $C=2.55$ for all curves. such fits are shown for different values of $t_{w}$ and $\Delta$ in Fig. \ref{fig: 5}b.

As shown earlier, the single-step relaxation of the elastic foams
can be described by a superposition of two logarithmic decays, offset by a time $t_{0}$ that depends on the compression. In analogy to Eq. 3, one can try to describe the non-monotonic behavior in the elastic foams using a superposition of four logarithmic
processes. Indeed, it can be shown that this introduces only small corrections
to the linear scaling between $t_{p}$ and $t_{w}$, in agreement with the data collapse in Fig. \ref{fig: 5}a. 
However, due to the nonlinearity introduced by the compression-dependent $t_{0}$, it is not possible to use the single-step
relaxations to obtain a good fit to the non-monotonic relaxation curve.

The non-monotonic relaxations reported here are reminiscent
of the aging behavior first described in the pioneering work of Kovacs
\cite{Kovacs1963}.  Kovacs examined the slow volume changes of polymer melts
following a temperature change, demonstrating memory retention in
a glassy system. Analogous phenomena was observed in the time-dependant
viscosity of metallic glasses \cite{Volkert1989} and density
of agitated granular systems \cite{Knight1995} as well as in numerical
studies \cite{Mossa2004,Cugliandolo2004}. Despite recent progress
\cite{Bertin2003,Prados2010,Diezemann2011,BouchbinderLanger2010,Prados2014}, this phenomenon is still not well understood.
Our observations, and their agreement with a phenomenological framework known to describe relaxation and aging in glassy systems is clear evidence that athermal mechanical systems can exhibit glassy dynamics and that the non-monotonic behavior described here may be generic to many disordered systems. Techniques such as direct visualization \cite{Aharoni2010,Cambou2011,thiria2011relaxation} and monitoring of acoustic emission \cite{houle1996acoustic,kramer1996universal} combined with a physical understanding of the systems used in this study \cite{witten2007stress,nelson1991polymerized,oppenheimer2015shapeable} may shed new light on the structural origin of the slow relaxation, non-monotonic aging and memory effects observed in these systems.

We thank F. Spaepen, Y. Oreg, Y. Imry, J. P. Bouchaud and T. A. Witten for illuminating
discussions. This work was supported by the National Science Foundation through the Harvard Materials
Research Science and Engineering Center (DMR-1420570). AA acknowledges support from the Milton
Fund. AA and SMR acknowledge support from the Alfred P. Sloan research foundation.

\end{document}